\documentclass[useAMS,usenatbib]{mnras}

\usepackage{graphicx}
\usepackage{amsmath}
\usepackage{natbib}
\usepackage{color}
\usepackage[T1]{fontenc}
\usepackage{aecompl}
\newcommand*\msolar{\mathop{}\!\,\mathcal{M}_{\odot}}

\newcommand*\kmps{\mathop{}\!\,\mathrm{km\,s}^{-1}}

\newcommand*\magnitude{\mathop{}\!\,\mathrm{mag}}
\newcommand*\vvmax{\mathop{}\!\,\mathrm{V}/\mathrm{V}_{\mathrm{max}}}
\newcommand*\meanvvmax{\mathop{}\!\,\left\langle \vvmax \right\rangle}
\newcommand*\vmax{\mathop{}\!\,\frac{1}{\mathrm{V}_{\mathrm{max}}}}

\title[Maximising Survey Volume with Voronoi Diagram]{Maximising Survey Volume for Large Area Multi-Epoch Surveys with Voronoi Tessellation}
\author[Marco C. Lam]{
Marco C. Lam$^{1}$\thanks{E-mail: mlam@roe.ac.uk}\\
\\
$^{1}$Institute for Astronomy, University of Edinburgh, Royal Observatory of Edinburgh, Blackford Hill, Edinburgh EH9 3HJ, UK\\
}

\begin{document}

\date{original form 2017 January 16}

\pagerange{\pageref{firstpage}--\pageref{lastpage}} \pubyear{2017}

\maketitle

\label{firstpage}

\begin{abstract}
The survey volume of a proper motion-limited sample is typically much smaller than a magnitude-limited sample. This is because of the noisy astrometric measurements from detectors that are not dedicated for astrometric missions. In order to apply an empirical completeness correction, existing works limit the survey depth to the shallower parts of the sky that hamper the maximum potential of a survey. The number of epoch of measurement is a discrete quantity that cannot be interpolated across the projected plane of observation, so that the survey properties change in discrete steps across the sky. This work proposes a method to dissect the survey into small parts with Voronoi tessellation using candidate objects as generating points, such that each part defines a `mini-survey' that has its own properties. Coupling with a maximum volume density estimator, the new method is demonstrated to be unbiased and recovered $\sim20\%$ more objects than the existing method in a mock catalogue of a white dwarf-only solar neighbourhood with Pan--STARRS\ 1-like characteristics. Towards the end of this work, we demonstrate one way to increase the tessellation resolution with artificial generating points, which would be useful for analysis of rare objects with small number counts.
\end{abstract}


\section{Introduction}

A number of types of transient, variable and moving sources are not rare, but their detection requires repeated observations of the same part of sky. This was not possible to perform over a large sky area until the era of digital astronomy. The highly automated observing runs and efficient digital detectors allow efficient data collection, while faster processors and the automated data reduction pipelines allow the production of high volume of output. Earliest attempts for such automations digitise the photographic plates from large sky area surveys, where measurements are made objectively with computer, as opposed to measuring manually. Large scale projects of this kind include the PPM catalogue~\citep{1991psc..book.....R, 1993psc..book.....B}, Automated Plate Machine Project~\citep{1992MNRAS.255..521E, 1995MNRAS.277..820E}, USNO\,A\,1.0, A\,2.0 and B\,1.0~\citep{1996AAS...188.5404M, 1998AAS...19312003M, 2003AJ....125..984M}, SuperCOSMOS~\citep{2001MNRAS.326.1279H, 2001MNRAS.326.1295H, 2001MNRAS.326.1315H}, UCAC\,1, 2, 3 and 4~\citep{2000AJ....120.2131Z, 2004AJ....127.3043Z, 2010AJ....139.2184Z, 2013AJ....145...44Z} and SUPERBLINK~\citep{2002AJ....124.1190L, 2003AJ....126..921L, 2005AJ....130.1247L, 2008AJ....135.2177L}, all of which had several epochs and simple tiling strategies.

In the current era of digital astronomy, some surveys continue to use simple tiling patterns where multiple pawprints are combined immediately to produce full coverage over a sky cell, for example in the UKIDSS~\citep{2007MNRAS.379.1599L}, four pawprints can cover a cell while VISTA employs six~\citep{2015A&A...575A..25S}. In other cases, SDSS Stripe 82 had nine epochs on average~\citep{2008MNRAS.386..887B}, ALLWISE has a coverage from 12 to over 200 frames~\citep{2014ApJ...783..122K, 2016ApJS..224...36K}; and the Pan-STARRS1~(PS1) 3$\pi$ Steradian Survey~(3SS) typically has 60 epochs~\citep{2013ApJS..205...20M}, The Dark Energy Survey will scan $\sim$5,000 deg$^2$ 10 times~\citep{2005astro.ph.10346T}, {\it Gaia} will have on average 81 transits, with over 140 in the most-visited parts of the sky at the end of the 5\,yr nominal mission~\citep{2016A&A...595A...1G}, and LSST will provide close to 1,000 epochs for half of the sky towards the end of the 10\,yr survey mission~\citep{2009arXiv0912.0201L}\footnote{https://github.com/LSSTScienceCollaborations/}. The key survey characteristics~(depth and epoch coverage) vary on small scales and in complex ways because tiling strategies/overlapping patterns for these surveys are extremely complicated to maximize coverage due to losses from, for example, chip gaps between CCDs and unfavourable observing conditions, unlike the situation with large-format photographic plates. This in turn complicates the analysis of any survey sample culled from them because optimal techniques like $V_{\mathrm{max}}$ require the precise survey characteristics. This problem becomes even more complex when different surveys are combined to expand the wavelength coverage and/or maximum epoch difference, for example when SDSS is combined with USNO-B\,1.0 to derive the proper motions~\citep{2004ApJS..152..103G, 2004AJ....127.3034M}, the survey typically has five epochs~(four from USNO-B and one from SDSS).

Existing methods tackle inhomogeneity by limiting the analysis to the shallower part of the survey and by applying a global correction in order not to run into unaccountable incompleteness~(e.g.\ \citealp[hereafter M17]{2006AJ....131..571H, 2017AJ....153...10M}). In view of this problem, when deriving the white dwarf luminosity function~(WDLF) with a maximum volume density estimator, \citet[hereafter RH11]{2011MNRAS.417...93R} measured the empirical photometric and astrometric uncertainty for different skycells as defined by the tiling of the photographic plates. \citet[hereafter LRH15]{2015MNRAS.450.4098L} improved the completeness correction due to kinematic selections. The methods described in these works allow an analysis to probe deeper, but in order to maximize the use of the data, we propose a new method based on Voronoi tessellation that can further maximize the analytical survey volume within which completeness and other biases can be corrected. Voronoi tessellation has been employed in defining simulation grids, clustering analysis, visualization etc. However, its property that partitions sources into well defined grids has not been used in all domains of astrophysics. By dividing the sky through Voronoi tessellation into a number of cells that is equal to the number of candidate sources, we can treat each cell as a mini-survey that has well-defined local properties. Analysis can also be performed at lower or higher resolution if needed.

In Section~\ref{sec:maths}, we will describe the mathematical framework and the construction of the simulated solar neighbourhood in Section~\ref{sec:mcsimulation}. The new method is applied to the simulated data in Section~\ref{sec:wdlf} under different selection criteria and we describe one procedure through which the resolution can be increased. The bias due to the choice of model is briefly discussed. In the final section, we discuss one possible way to increase the resolution for the analysis and conclude this work.

\section{Mathematical Framework of the Voronoi Method}
\label{sec:maths}
The maximum volume density estimator~\citep{1968ApJ...151..393S} tests the observability of a source by finding the maximum volume in which it can be observed by a survey~(e.g.\ at a different part of the sky at a different distance). It is proven to be unbiased~\citep{1976ApJ...207..700F} and easily can combine multiple surveys~\citep{1980ApJ...235..694A}. In a sample of proper motion sources, we need to consider both the photometric and astrometric properties~(see LHR15 for details). The number density is found by summing the number of sources weighted by the inverse of the maximum volumes. For surveys with small variations in quality from field to field and from epoch to epoch, or with small survey footprint areas, the survey limits can be defined easily. However, in modern surveys, the variations are not small; this is especially true for ground-based observations. Therefore, properties have to be found locally to analyse the data most accurately. Through the use of Voronoi tessellation, sources can be partitioned into individual 2D cells within which we assume the sky properties are defined by the governing source. Each of these cells has a different area depending on the projected density of the population.

An important assumption for using the Voronoi method is that the distributions of the observing parameters of the cells at different resolutions are very similar to each other, hence the integrated maximum volume is approximately equal to the exact solution. In the rest of the article, {\it cell} will be used to denote Voronoi cell and {\it h-pixel} for HEALPix pixel~\citep[see Section~\ref{sec:volume_new}]{2005ApJ...622..759G} and {\it pixels} for the ones on a detector.

\subsection{Voronoi Tessellation}

A Voronoi tessellation is made by partitioning a plane with $n$ points into $n$ convex polygons such that each polygon contains one point. Any position in a given polygon~(cell) is closer to its generating point than to any other for the case of Voronoi tessellation using Euclidean distance~$\left( D_E = \sqrt{(x_1-x_2)^2 + (y_1-y_2)^2} \right)$. For use in astronomy, such a tessellation has to be done on a spherical surface~(two-sphere).

In the following work, the tessellation is constructed with the {\sc SciPy} package spatial.SphericalVoronoi, where each polygon is given a unique ID that is combined with the vertices to form a dictionary. The areas are calculated by first decomposing the polygons into spherical triangles with the generating points and their vertices\footnote{https://github.com/tylerjereddy/py\_sphere\_Voronoi} and then by using L'Huilier's Theorem to find the spherical excess. For a unit-sphere, the spherical excess is equal to the solid angle of the triangle. The sum of the constituent spherical triangles provides the solid angle of each cell.

\subsection{Cell Properties}

For a Voronoi cell $j$, the properties of the cell are assumed to be represented by generating source $i$. Both $i$ and $j$ are indexed from $1$ to $\mathcal{N}$, but since each source has to be tested for observability in each cell to calculate the maximum volume, $i$ and $j$ cannot be contracted to a single index. Furthermore, the cells do not need to be defined by only the sources. Arbitrary points can be used for tessellation such that $i$ and $j$ will not have a one-to-one mapping. The epoch of the measurement is labelled by $k$. When tested for observability, epoch-wise information is essential in calculating the photometric and proper motion uncertainties as functions of distance. The major difference in the following approach is that the proper motion uncertainty is found from the formal propagation of errors instead of measuring the empirical form as a function of magnitude, $\sigma_\mu($mag$)$, which limits the survey to the worst part of a tile~(RH11). This new approach does not need to take into account the scatter in $\sigma_\mu($mag$)$, due to different local sky properties and different colours of the sources and their neighbours. Different types of source can differ by up to a few magnitudes in the optical/infrared colours, so two sources with similar magnitudes in one filter can have very different proper motion uncertainties if one is close to the detection limit in another filter. The modelling of the photometric uncertainties from CCD detectors is much simpler than that for photographic plates, because the photometric response of the modern detectors is much more linear at both the faint and bright ends. Thus, the uncertainties can be estimated with relatively simple equations.

\subsubsection{Photometric Uncertainty}
\label{sec:photometric_uncertainty}
When a source is being tested for the observability, it is `placed' at a different distance so the apparent brightness changes as a consequence. The background and other instrumental noises are constant, but the Poisson noise from the source changes with the measured flux, hence the photometric uncertainties are functions of distance. The total noise, $N$, of a photometric measurement can be estimated by 
\begin{equation}
\label{eq:photometric_noise}
N = \sqrt{ ( F + d + s ) \times t + r^2 }
\end{equation}
where $F$ is the instrumental flux per unit time, $d$ is the dark current per unit time, $s$ is the sky background flux per unit time, $t$ is the exposure time and $r$ is the read noise. Among these quantities, $d$, $s$, $t$ and $r$ are fixed quantities in a given epoch, only the flux varies as a function of distance. We use $F$ as the measured flux and $\mathcal{F}(D)$ as the flux at an arbitrary distance $D$. Therefore, in a Voronoi cell $j$ at epoch $k$, the photometric noise of source $i$ is
\begin{equation}
N_{i,j,k}(D) = \sqrt{ \left( \mathcal{F}_{i}(D) + d_{j,k} + s_{j,k} \right) \times t_{j,k} + r^2 }
\end{equation}
where the flux at $D$ is calculated from applying the inverse square law on the observed flux $F_i$ and observed distance $D_i$,
\begin{equation}
\mathcal{F}_{i}(D) = F_{i} \times \left( \dfrac{D_{i}}{D} \right)^2 .
\end{equation}
The random photometric uncertainty of a source at an arbitrary distance in a given epoch is the inverse signal-to-noise ratio,
\begin{equation}
\delta \mathcal{F}_{i,j,k}(D) = \dfrac{N_{i,j,k}(D)}{\mathcal{F}_{i}(D)}.
\end{equation}
The total photometric uncertainty of the source as a function of distance, combining with the systematic uncertainty, $\sigma_s$, coming from the absolute calibration of the detector, is therefore
\begin{equation}
\sigma_{i,j,k}(D) = \sqrt{ \delta \mathcal{F}^2_{i,j,k}(D) + \sigma^2_{s} },
\end{equation}
which represents the photometric uncertainty as a function of the distance to the source.

\subsubsection{Astrometric Uncertainty}
The least square solution of proper motion in one direction for source $i$ can be expressed in the following matrix form, the epoch is labelled by the subscript from $1$ to $M(j)$ where $M$ is the number of epochs in cell $j$,
\begin{equation}
\underbrace{\left(
\begin{array}{cc}
\dfrac{1}{\sigma_{1}} & \dfrac{\Delta t_{1}}{\sigma_{1}} \\
\cdot & \cdot  \\
\cdot & \cdot  \\
\dfrac{1}{\sigma_{M(j)}} & \dfrac{\Delta t_{M(j)}}{\sigma_{M(j)}}
\end{array}
\right)}_{\mathbf{A}}
\times
\left(
\begin{array}{c}
\overline{\alpha}\\
\mu_{\alpha}
\end{array}
\right)
=
\left(
\begin{array}{c}
\dfrac{\Delta\alpha_{1}}{\sigma_{1}}\\
\cdot\\
\cdot\\
\dfrac{\Delta\alpha_{M(j)}}{\sigma_{M(j)}}
\end{array}
\right)
\end{equation}
where $\Delta t_k$ is the time difference between the mean epoch and epoch $k$, $\Delta \alpha_{k}$ is the positional offset from the mean position, $\overline{\alpha}$, and proper motion, $\mu_{\alpha}$, in the direction of the right ascension. The associated uncertainties can be found from the diagonal terms of the normal matrix,
\begin{equation}
\mathbf{\mathsf{A}}^{\mathrm{T}}\mathbf{\mathsf{A}}
=
\left[
\begin{array}{cc}
\sum\limits_{k} \left( \dfrac{1}{\sigma_{k}} \right)^2 & \sum\limits_{k} \left( \dfrac{1}{\sigma_{k}} \dfrac{\Delta t_k}{\sigma_{k}} \right) \vspace*{1mm}\\
\sum\limits_{k} \left( \dfrac{1}{\sigma_{k}} \dfrac{\Delta t_k}{\sigma_{k}} \right) & \sum\limits_{k} \left( \dfrac{\Delta t_k}{\sigma_{k}} \right)^2
\end{array}
\right]
\end{equation}
so for each cell,
\begin{equation}
\dfrac{1}{\sigma_{\mu_{\alpha}\cos\delta}^{2}} = \sum_{k} \left( \dfrac{\Delta t_k}{\sigma_{k}} \right)^2
\end{equation}
and the total proper motion uncertainty is
\begin{equation}
\label{eq:sigma_mu}
\sigma_{\mu} = \sqrt{ \sigma^2_{\mu_{\alpha\cos\delta}} + \sigma^2_{\mu_{\delta}} } = \sqrt{2} \sigma_{\mu_{\alpha\cos\delta}} .
\end{equation}
The uncertainties in the $\alpha$ and $\delta$ directions are symmetrical in four-parameter astrometric solution~(two positions and two proper motions). In the case of five-parameter solution where parallax is solved and for the seven-parameter solution where, in addition, the acceleration terms in both directions are solved for, the uncertainties will not be symmetrical due to the parallactic term, so the off-diagonal terms have to be taken into account which would otherwise be negligible compared to the diagonal terms. However, for a variance-weighted mean epoch, the off-diagonal terms are exactly zero by definition.

\subsection{Consequence to $V_{\mathrm{max}}$ Calculation}
\label{sec:volume_new}
There is only one minor adjustment to the volume integral -- the lower proper motion limit. Instead of finding the limit by measuring from a number of nearby sources that include mostly sources with different colours, the limit is defined by the properties of the Voronoi cell that comes only from the generating source of the cell. A different set of pixelization by HEALPix, denoted by $l$, is for the line-of-sight tangential velocity completeness correction~(see RH11 for detailed description). For source $i$, the maximum volume has to be tested in each cell $j$, the expression is almost identical to that in LRH15, except for the $j$ and $l$ indexes
\begin{multline}
\label{eq:vmax}
\mathrm{V}_{\mathrm{max}} = \sum_j \Omega_j \int_{D_{\mathrm{min,j}}}^{D_{\mathrm{max,j}}} \dfrac{\rho(D)}{\rho_{\odot}} \times D^2 \\
\hfill \times \left[ \int_{a(D)}^{b(D)} P_{l(j)}(v_{\mathrm{T}}) \, \mathrm{d}v_{\mathrm{T}} \right] \mathrm{d}D
\end{multline}
where $\dfrac{\rho(D)}{\rho_{\odot}}$ is the density normalized by that at the solar neighbourhood, $P_{l(j)}$ is the tangential velocity distribution, $l(j)$ denotes the h-pixel mapped from cell $j$ with area $\Omega_j$, $v_{\mathrm{T}}$ is the tangential velocity, $D_{\mathrm{min}}$ and $D_{\mathrm{max}}$ are the minimum and maximum photometric distances, and $\sigma_{\mu}(D)$ is the proper motion uncertainty as a function of the distance to the source. This is to model the change in the proper motion uncertainties due to varying apparent magnitude with distance~(i.e.\ at greater distance the proper motion uncertainty will be larger because the source becomes fainter which increases the single-epoch positional uncertainty). The lower tangential velocity limit in the inner integral, $a(D)$, is
\begin{equation}
\label{eq:inner_limit}
a(D) = \mathrm{max}\left[v_{\mathrm{min}}, 4.74\times s \times\sigma_{\mu}(D) \times D \right]
\end{equation}
where the factor of 4.74 comes from the unit conversion from arcsec yr$^{-1}$ to km s$^{-1}$ at distance $D$, $v_{\mathrm{min}}$ is the global lower tangential velocity limit and $s$ is the significance of the proper motions. The expression is identical to that in LRH15, but $\sigma_{\mu}(D)$ is calculated in a completely different way.

The inner integral can vanish before reaching the distance limits so the integrator must use a small step size or the distances at which the inner integral vanish have to be calculated explicitly. The cell ID $j$ and h-pixel ID $l$ can be set as a one-to-one mapping by calculating the tangential velocity distribution for each of the Voronoi cell. However, the Voronoi tessellation is dependent on the sample, while the tangential velocity correction is fixed on the sky, using a precomputed look-up table because the tangential velocity correction can significantly reduce computation time.

\section{Simulated Data Set}
\label{sec:mcsimulation}
To demonstrate the power of the Voronoi method described in Section~\ref{sec:maths}, we apply it to catalogues of simulations of the solar neighbourhood. This section details the construction of the Monte Carlo simulation.

We generated snapshots of white dwarf (WD)-only populations in the solar neighbourhood containing six dimensional phase space information. The procedure is very similar to that described in LRH15, however, we introduce changes to the noise model of the system and include epoch-wise information. The volume probed is assumed to be small such that the simulation is performed in a Cartesian space, instead of a plane polar system centred at the Galactic Centre. The Galaxy has three distinct kinematic components: a thin disc, a thick disc and a stellar halo, all of which we model with no density variations along the co-planar directions of the Galactic plane. The vertical structures of the discs follow exponential profiles, with scale height $H_{\mathrm{thin}}$ and $H_{\mathrm{thick}}$ such that
\begin{equation}
\dfrac{\rho(D)}{\rho_{\odot}} = \exp\left(-\dfrac{\left\vert z \right\vert}{H} \right) = \exp\left(- \dfrac{\left\vert D \sin b \right\vert}{H} \right),
\end{equation}
where $z$ is the vertical distance from the Galactic plane and $b$ the Galactic latitude. None of the three components are tilted relative to each other. The velocity components, $U$, $V$ and $W$, of each WD are drawn from the Gaussian distributions described by the measured means and standard deviations of the three sets of kinematics that describe the three populations in the solar neighbourhood~(Table~\ref{table:parameters}). 

Theoretical WDLFs are used as the probability distribution functions~(pdfs) in the simulations. The normalizations of the pdfs are taken from the WD densities found in RH11. The input parameters for a WDLF are the star formation rate~(SFR), initial mass function~(IMF), MS evolution model and WD cooling model. The standard equation for modelling the WDLF with those four given inputs is
\begin{equation}
\Phi(M_{\mathrm{bol}}) = \int_{\mathcal{M}_{l}}^{\mathcal{M}_{u}} \frac{\mathrm{d}t_{\mathrm{cool}}}{\mathrm{d}M_{\mathrm{bol}}} \psi \left( t_0 - t_{\mathrm{cool}} - t_{\mathrm{MS}} \right) \phi\left( \mathcal{M} \right) \mathrm{d}\mathcal{M},
\end{equation}
where $\Phi(M_{\mathrm{bol}})$ is the number density of WDs at magnitude $M_{\mathrm{bol}}$. The derivative inside the integral is the characteristic cooling time of WDs, $\psi(t)$ is the SFR at time $t$ and $\phi$ is the IMF.
The input parameters are assumed to be invariant with time and are summarized in Table~\ref{table:parameters}. The integral also depends on the lifetimes of MS progenitors, $t_{\mathrm{MS}}$, as a function of mass and metallicity. We have adopted the stellar evolution tracks from the Padova group~(PARSEC;~\citealt{2012MNRAS.427..127B}) with a metallicity of $Z=0.019$ and $Y=0.30$~\citep{2000A&AS..141..371G}. By assuming a fixed surface gravity $\log{g}=8.0$ and pure hydrogen atmosphere~(DA)\footnote{http://www.astro.umontreal.ca/$\sim$bergeron/CoolingModels/}, the WD cooling time, $t_{\mathrm{cool}}$, is a function of mass and luminosity~\citep{2006AJ....132.1221H, 2006ApJ...651L.137K, 2011ApJ...737...28B, 2011ApJ...730..128T}, and $t_0$ is the total time since the onset of star formation. The integral is over all MS masses that have had time to produce WDs at the present day, with the magnitude-dependent lower limit, $\mathcal{M}_{l}$, corresponding to the solution of
\begin{equation}
t_0 = t_{\mathrm{cool}} \left( M_{\mathrm{bol}},\xi(\mathcal{M}_{l}) \right) + t_{\mathrm{MS}} \left( \mathcal{M}_l, Z \right)
\end{equation}
and the upper limit for WD production $\mathcal{M}_u \approx 8 \mathcal{M}_{\odot}$. The initial--final mass relation, $\xi$, relates the MS progenitor mass to the mass of the WD, is adopted from \citet{2009ApJ...705..408K} without including globular clusters in the analysis where the final WD mass can be expressed as
\begin{align}
\xi\left(\mathcal{M}_{i}\right) = \mathcal{M}_{f}\left(\mathcal{M}_{i}\right) = 0.101\mathcal{M}_{i} + 0.463\msolar.
\end{align}
The thin and thick disc populations are assigned with constant SFR since look back time, $\tau = 8$\,Gyr and $\tau=10$\,Gyr respectively, while the halo has a starburst of $1$\,Gyr at $\tau=13$\,Gyr. The IMF has an exponent of $-2.3$ in the mass range of interest~\citep{2001MNRAS.322..231K}.

\begin{table}
\centering
\caption{Physical properties of the Galaxy used in the Monte Carlo simulation.}
\label{table:parameters}
\begin{tabular}{ l *{3}{c} }
\hline
\hline
Parameter & Thin disc & Thick disc & Stellar halo\\
\hline
$\langle\mathrm{U}\rangle$/km s$^{-1}$ & $-8.62^{a}$ & $-11.0^{d}$ & $-26.0^{d}$ \\
$\langle\mathrm{V}\rangle$/km s$^{-1}$ & $-20.04^{a}$ & $-42.0^{d}$ & $-199.0^{d}$ \\
$\langle\mathrm{W}\rangle$/km s$^{-1}$ & $-7.10^{a}$ & $-12.0^{d}$ & $-12.0^{d}$ \\
$\sigma_{\mathrm{U}}$/km s$^{-1}$ & $32.4^{a}$ & $50.0^{d}$ & $141.0^{d}$ \\
$\sigma_{\mathrm{V}}$/km s$^{-1}$ & $23.0^{a}$ & $56.0^{d}$ & $106.0^{d}$ \\
$\sigma_{\mathrm{W}}$/km s$^{-1}$ & $18.1^{a}$ & $34.0^{d}$ & $94.0^{d}$ \\
H/pc & 250$^{b}$ & 1000$^{e}$ & $\infty$ \\
n/pc$^{-3}$ & 0.00310$^{c}$& 0.00064$^{c}$& 0.00019$^{c}$ \\
\hline
\end{tabular}
\begin{enumerate}
\itemsep0em
\item[a] \citealt{2009AJ....137..266F}.
\item[b] \citealt{1998A&A...333..106M}.
\item[c] RH11.
\item[d] \citealt{2000AJ....119.2843C}.
\item[e] \citealt{1987AJ.....93...74S}.
\end{enumerate}
\end{table}

From the pdfs of the kinematics, distance and bolometric magnitude, we calculate the apparent magnitudes in the PS1 g$_{\rm P1}$, r$_{\rm P1}$, i$_{\rm P1}$, z$_{\rm P1}$ and y$_{\rm P1}$ filters\footnote{http://panstarrs.stsci.edu/}~\citep{2012ApJ...750...99T, 2012ApJ...756..158S, 2013ApJS..205...20M, 2016arXiv161205560C, 2016arXiv161205245W, 2016arXiv161205244M, 2016arXiv161205243F, 2016arXiv161205242M, 2016arXiv161205240M}. The line-of-sight and the projected velocity can be derived from the given 3D kinematics and 3D position. The uncertainties in the five filters are calculated from the sky background flux, exposure time, dark current and read noise that are representative of the PS1 3SS at Processing Version~2~(PV2). The sky background flux is drawn from a Gaussian distribution measured from the 3SS in each of the filters~(Table \ref{table:noise_model}). The means and standard deviations\footnote{$1.4826$ times the median absolute deviation is used for robust estimation of the standard deviation.} were measured from $100$ fields drawn randomly across the survey footprint area.

To simulate the variations in the observing properties of the sky, HEALPix is used to pixelate the sky using a resolution of $n_{\mathrm{side}} = 256$, i.e.\,each has a size of $\sim0.0534$ deg$^{2}$, which is sufficiently small compared to the projected density of white dwarfs that is less than $1$ deg$^{-2}$~(e.g.\ RH11). The HEALPix resolution is on a much finer scale than the Voronoi tessellation used for determining the maximum volume in order to test the accuracy later~(Section~\ref{sec:resolution}). One feature of this approach is that when the analysis is done at a higher resolution, the new set of cells will be guaranteed to land on a different set of h-pixels and as such will provide a self-consistency check. There is not a switching from Voronoi cells to HEALPix pixels in the analysis: there is simply a look-up table for matching a given cell with the h-pixel that the cell-generating source lands on. Each h-pixel is given a sky background noise for each epoch of the measurement. When the sky brightness is below the lower limit, a sky brightness that is fainter than $99.73\%$ of the measured values, the background noise is resampled until it is above the limit. An ADU of $1$ e$^{-}$ per photon is assumed. The 3SS has $12$ epochs on average in each of the filters, so the number of epochs for each source in the simulation is drawn from a distribution\footnote{This study only focuses on tessellation; the effect of non-detection is another huge step in the optimization of analysis.} that follows $1+P(11)$, where $P(11)$ is Poisson distribution with a mean of $11$ and the epochs are drawn from a random distribution over a period of $3$\,yr with $6$\,h on either side of the source masked out to simulate seasonal observing. When sources are distributed over the sky, they will take the set of values defined by the nearest pixel. Using the treatments from Section~\ref{sec:photometric_uncertainty}, with a dark current of $0.2\,$e$^{-}$s$^{-1}$, exposure times of $43, 40, 35, 30$ and $30$ s, zero-point magnitudes at $24.563$, $24.750$, $24.611$, $24.250$ and $23.320$\,mag in the five filters, and a constant read noise of $5.5\,$e$^{-}$~\citep{2013MNRAS.435.1825M}, each source is assigned with proper motion uncertainty using equation~(\ref{eq:sigma_mu}). These inputs produce an all-sky survey that has $10\sigma$ detections~(and their standard deviations) in g$_{\rm P1}$, r$_{\rm P1}$, i$_{\rm P1}$, z$_{\rm P1}$ and y$_{\rm P1}$ at $21.98 \pm 0.04$, $21.53 \pm 0.05$, $21.12 \pm 0.04$, $20.54 \pm 0.05$ and $19.59 \pm 0.04\magnitude$. They are similar to the PV2 values, but the distribution is much narrower because the noise model is itself noiseless~(e.g.\ the sources are not affected by diffraction spikes, optical ghosts, cosmic rays or other effects that lead to larger scatter).

\begin{table}
\centering
\caption{Parameters for the sky background count.}
\label{table:noise_model}
\begin{tabular}{ *{4}{c} }
\hline
\hline
Filter & Mean & Standard Deviation & Lower limit \\
 & (photon s$^{-1}$) & (photon s$^{-1}$) & (photon s$^{-1}$) \\
\hline
g & $39.972$ & $12.355$ & $16.761$ \\
r & $135.728$ & $47.981$ & $50.712$ \\
i & $262.725$ & $72.499$ & $108.612$ \\
z & $257.930$ & $90.175$ & $92.511$ \\
y & $272.195$ & $72.493$ & $88.358$ \\
\hline
\end{tabular}
\end{table}

\section{Application to WDLFs}
\label{sec:wdlf}
This section describes the application of our survey dissection method~(presented in Section~\ref{sec:maths}) to simulated PS1-like WD catalogues generated using the recipe described in Section~\ref{sec:mcsimulation}. The bright limits at all filters are set at $15\magnitude$. The faint limits are at $21.5$, $21.0$, $20.5$, $20.0$ and $19.5\magnitude$ in g$_{\rm P1}$, r$_{\rm P1}$, i$_{\rm P1}$, z$_{\rm P1}$ and y$_{\rm P1}$ filters respectively, which are the typical magnitudes at which the 3SS is complete. The lower proper motion limit is set to five times the proper motion uncertainties, $\sigma_\mu$, unless specified otherwise; and the upper proper motion limit is set at $0.08438$ and $0.4219$\,arcsec\,yr$^{-1}$ for the cases using lower tangential velocity limits at $40$ and $200\kmps$, respectively. The two limits correspond to a minimum distance of $100$\,pc; this is to avoid any bias coming from the unaccounted parallax signature of very nearby sources. The upper tangential velocity limits are different in each analysis. The photometric parallaxes were not derived, and the real distances and bolometric magnitudes are used. The volume and the maximum volume are found by integrating equation~(\ref{eq:vmax}) from $D_\mathrm{min}$ to $D$, and from $D_\mathrm{min}$ to $D_\mathrm{max}$ respectively.

\subsection{Comparison with the RH11 selection}

\begin{figure}
\includegraphics[width=\columnwidth]{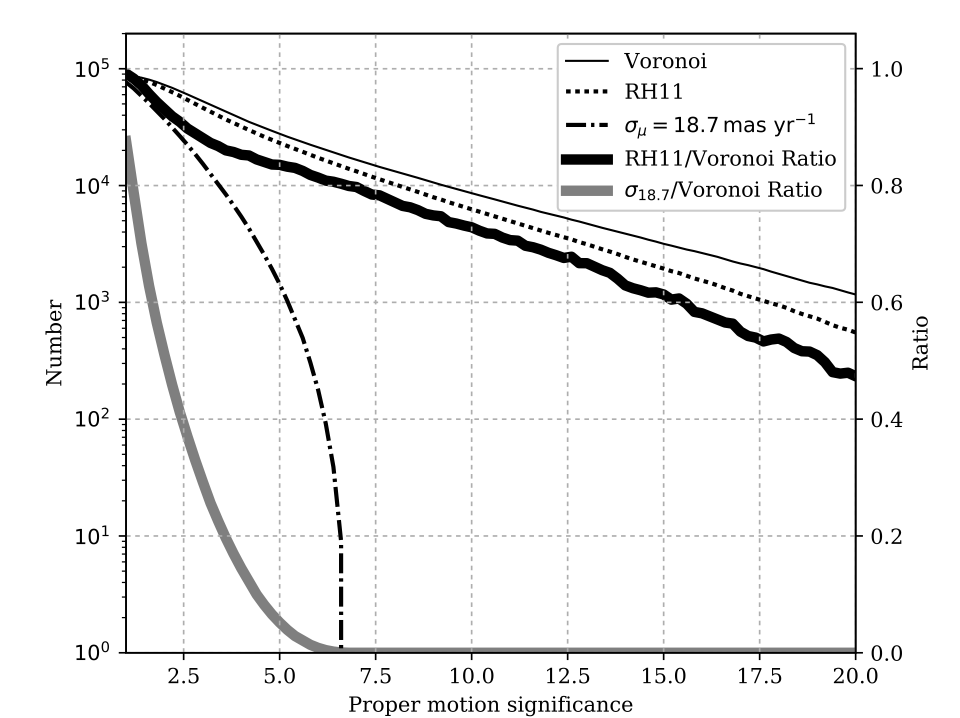}
\caption{The number of sources recovered using the Voronoi and RH11 methods in a mixed thin disc, thick disc and halo simulation as a function of proper motion significance. The Voronoi method is plotted as a thin solid line, RH11 method is plotted as a dashed line, values correspond to the ordinate axis in logarithmic scale on the left. The use of a global proper motion uncertainty at $95$th percentile~(i.e.\ $18.7\ \mathrm{mas}\ \mathrm{yr}^{-1}$) instead of RH11 tiling is shown by the dot--dashed line. The ratios between the two methods to the Voronoi method are plotted as thick solid lines and correspond to the ordinate axis on the right.}
\label{fig:rh11_pm_sig}
\end{figure}

\begin{figure}
\includegraphics[width=\columnwidth]{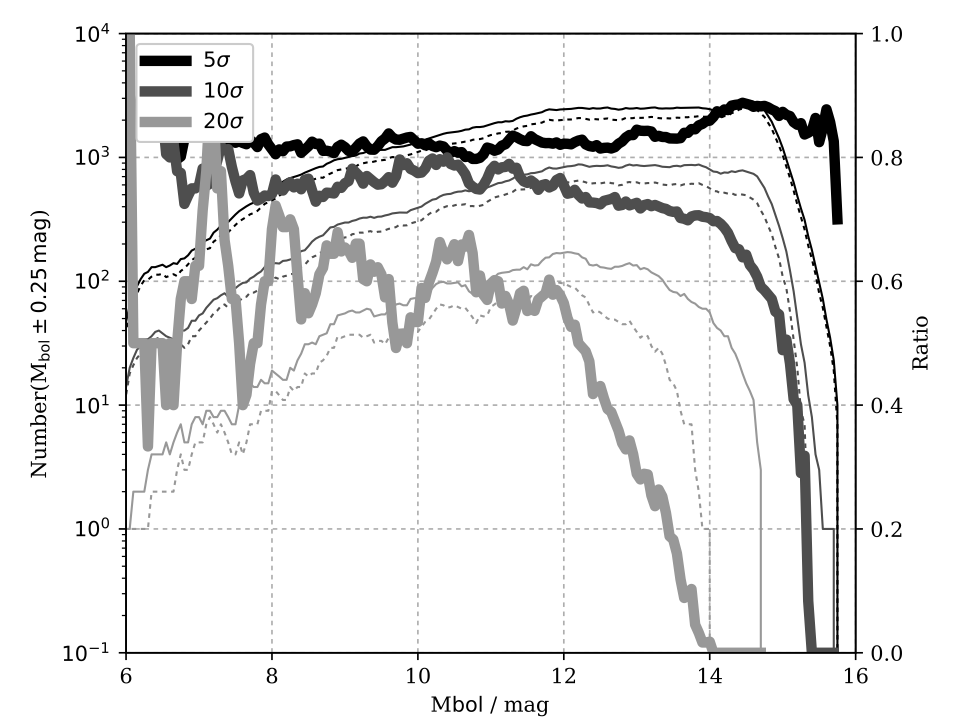}
\caption{The number of sources in a moving $0.5\magnitude$ bin discovered as a function of absolute bolometric magnitude with the Voronoi method~(solid) and RH11 method~(dashed) at $5\sigma$~(black), $10\sigma$~(grey) and $20\sigma$~(light grey) level, values correspond to the ordinate axis in logarithmic scale on the left. The ratios between the two methods are in thick solid lines and correspond to the ordinate axis on the right.}
\label{fig:rh11_mbol}
\end{figure}

The RH11 method increases the survey volume by restricting shallow survey depths only over areas that are severely limited by a small number of poor observations. In this section we illustrate how the Voronoi method can further increase the number of sources that can be recovered while rigorous completeness correction can be performed. In Fig.~\ref{fig:rh11_pm_sig}, under the selection criteria: $40 \kmps < v_{\mathrm{tan}} < 60 \kmps$, proper motion less than half an arcsec\,year$^{-1}$ and a minimum distance of $100$\,pc, the number of sources recovered by the Voronoi method is plotted as a solid line, RH11 method as a dashed line and the global $95$th percentile as a dotted line. The ratio between the RH11 and Voronoi methods is plotted as a thick black solid line, while that between a global lower proper motion limit and the Voronoi method is plotted as a thick grey solid line. With the Voronoi method, more sources can be recovered. The ratio between the RH11 and Voronoi methods slowly decreases as the absolute bolometric magnitudes increase; the ratio with the global limit simply plummets given that the upper proper motion limit is only $0.5$\,arcsec\,yr$^{-1}$. Fig.~\ref{fig:rh11_mbol} shows that the ratios stay fairly constant until some faint limits. The sources lost with the RH11 treatment are due to a combination of (1)~the reassignment of proper motion uncertainties based on empirical observations where $95$-per-cent of all sources are given larger uncertainty values than their measured ones and (2)~the loss of the deepest areas of the survey. 

Due to the simplistic noise model of the simulation~(i.e.\ no bad pixels, saturation, diffraction, optical ghosts and other effects that affect the photometric and astrometric precision significantly), the distribution of the proper motion uncertainties in the simulation is typically narrower than real measurements. Nevertheless, at $5\sigma$ level the Voronoi method can recover $\sim15\%$ more sources than the RH11 method and much more sources than applying a global lower proper motion limit~(Fig.~\ref{fig:rh11_mbol}).

\subsection{Thin Disc and Combined Discs}

Study of the thin disc WDLF requires a selection of the low-velocity population in order to minimize the contaminations from older populations, which typically possess higher velocities. In this section, we show the observed WDLFs from a thin disc-only simulation and from a mixed thin disc, thick disc and halo simulation. The WDLF comparison plots are displayed with the WDLF in the top panel, differences between the input and calculated WDLFs and the $\meanvvmax$ as a function of bolometric magnitude are in the middle panel and at the bottom panel respectively.

\subsubsection{Thin disc-only sample}

\begin{figure}
\includegraphics[width=\columnwidth]{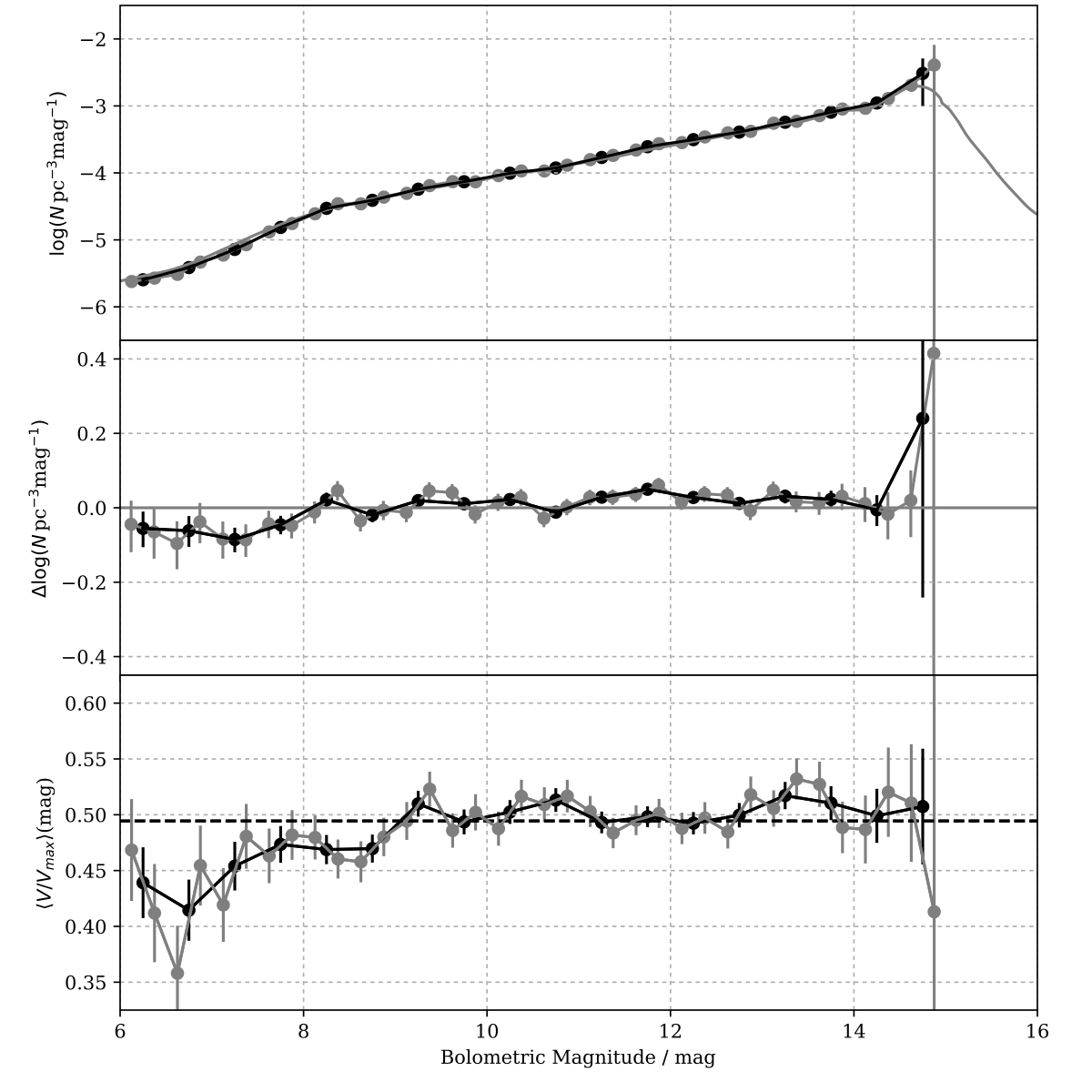}
\caption{Top: the WLDFs for a thin disc-only realization using the new method, where $0.5$ and $0.25\magnitude$ binning are displayed in black and grey, respectively.  The light grey line shows the input function. Middle: the deviations of the WDLF as a function of absolute bolometric magnitude. Bottom: the $\meanvvmax$ as a function of absolute bolometric magnitude. The dashed line indicates the overall $\meanvvmax$.}
\label{fig:thin_wdlf}
\end{figure}

In an analysis selecting only thin-disc WDs, the observed WDLF agrees very closely with the input function down to $M_{\mathrm{bol}} \approx15\magnitude$ when the number of sources drops significantly~(Fig.~\ref{fig:thin_wdlf}). From the $\meanvvmax(\mathrm{mag})$ distribution, the derived solution is very stable throughout, except at the brightest and the faintest ends where the $\vmax$ method is known to become less reliable as the number of sources decreases. In theory, $\meanvvmax = 0.5$, because it is the expected value of a uniform distribution between $0$ and $1$. Statistically, it is expected that only $\sim60\%$ of the time the $\meanvvmax$ lies within the error bar. The uncertainty in $\meanvvmax$ is $\frac{1}{\sqrt{12N}}$. The small oscillation about the line at $\meanvvmax(\mathrm{mag})=0.5$ is a good indication that the sample is unbiased over a large dynamic range of magnitudes. The outliers at the extreme ends result from the application of the density estimator to a small number of sources, and so likely do not represent the true values. Taking $40$ and $60\kmps$ as the lower and upper tangential velocity limits of the inner integral~(equation~\ref{eq:vmax} \& \ref{eq:inner_limit} and the equivalent set of the upper limit), the total integrated number density of the work is $3.65 \pm 1.03 \times 10^{-3}$ pc$^{-3}$, compared to the input $3.10 \times 10^{-3}$ pc$^{-3}$. The overall $\meanvvmax = 0.4944 \pm 0.0031$, which is very close to $0.5$, indicating an unbiased sample.

\subsubsection{Mixed Population~($40-60\kmps$)}

The modification to the density estimator itself is small, only the lower limit of the inner integral is changed~(equation~\ref{eq:inner_limit}); and it is not expected that the effect due to contamination should differ from the previous analysis in LRH15. The extra depth enabled by the new method could have led to a significant increase in the measured density due to a combination of two effects: (1)~an increase in contamination fraction as the thin disc contribution drops rapidly with distance: at the Galactic poles, the thin disc and thick disc densities equate at $525$\,pc; and (2)~the kinematic completeness correction applied on contaminants, which are more common at fainter magnitudes.

\begin{figure}
\includegraphics[width=\columnwidth]{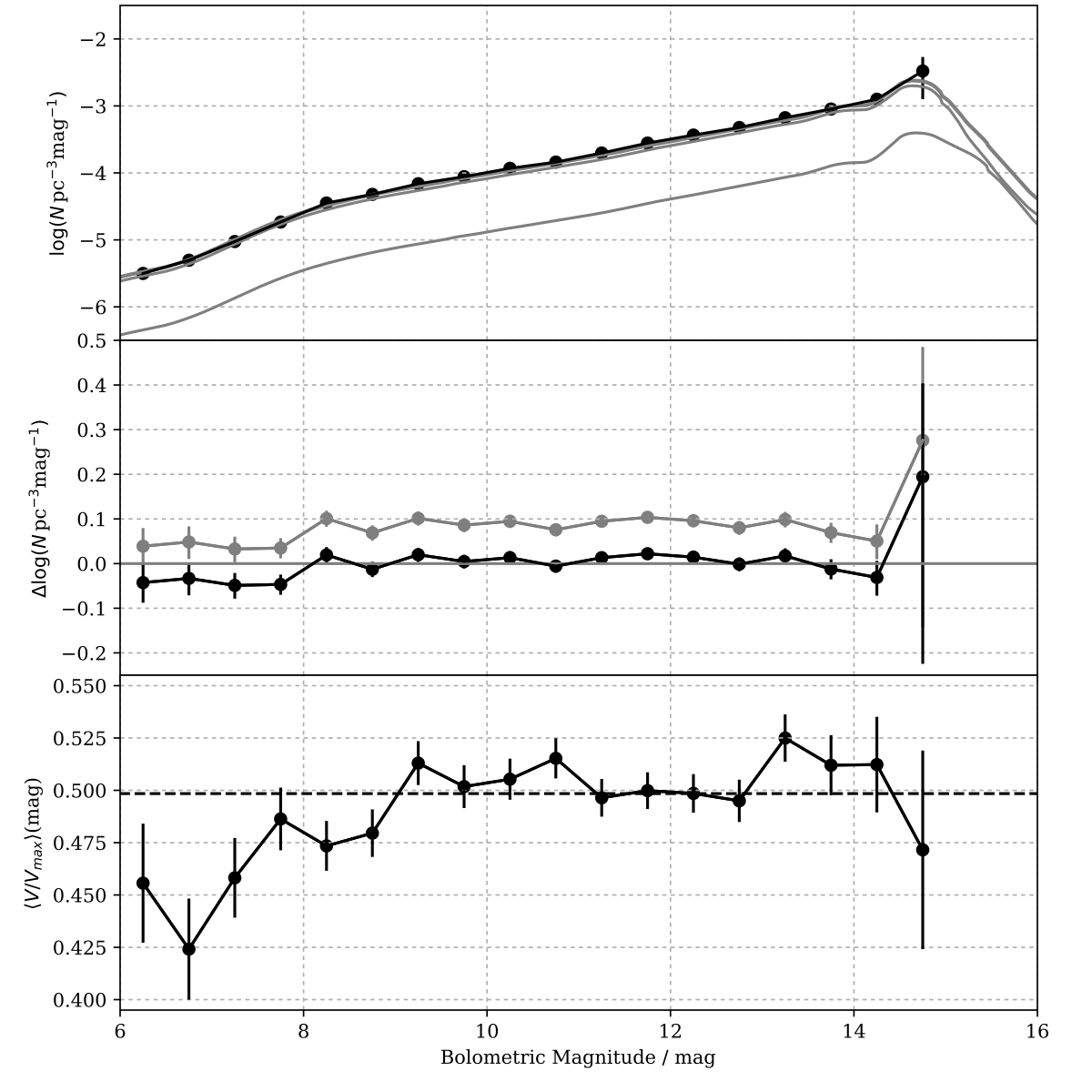}
\caption{Top: WLDF for all sources with tangential velocities between $40$ and $60\kmps$ from a mixed-population catalogue using the new method assuming thin disc properties bins~(black), the input thin and thick disc luminosity functions~(thin grey line) and the combined thin and thick disc luminosity function~(thick grey line). Middle: the deviations of the WDLF as a function of absolute bolometric magnitude from the thin disc luminosity function~(grey) and from the combined luminosity function~(black). Bottom: the $\meanvvmax$ as a function of absolute bolometric magnitude. The dashed line indicates the overall $\meanvvmax$.}
\label{fig:40_60_wdlf}
\end{figure}

The kinematics of the two discs are well measured; however, the relative density of WDs in them is much less studied -- there is only one measurement on record~(RH11). To understand the effects of contamination, a better understanding of the two populations is needed. Nevertheless, we can compare the WDLFs from the last section to a mixed population with the same set of upper and lower tangential velocity limits~($40$ and $60\kmps$; Fig.~\ref{fig:40_60_wdlf}). The total integrated number density is $4.00 \pm 1.03 \times 10^{-3}$ pc$^{-3}$ as compared to $3.10 \times 10^{-3}$ pc$^{-3}$ for the thin disc and $0.64 \times 10^{-3}$ pc$^{-3}$ for the thick disc, which sum to $3.74 \times 10^{-3}$ pc$^{-3}$. If it is treated as a pure thin disc WDLF, there is a roughly constant overestimation of $0.1$\,dex at all magnitudes. When both discs are considered, the small over-density~($0.26 \times 10^{-3}$ pc$^{-3}$) is due to contamination from the thick disc where the using of a thin disc scale-height on these sources will lead to an overestimation of the maximum volume. In this simulation, $16.0\%$ of the data are from the thick disc.

The $\meanvvmax$ distribution is very similar to the clean sample, and for the entire sample $\meanvvmax = 0.4984 \pm 0.0028$ which is within 1 standard deviation of the ideal value $0.5$. We believe this velocity range is a good choice for driving an upper limit of the thin disc white dwarf density in the solar neighbourhood.

\subsubsection{Mixed Population~($40-120\kmps$)}

\begin{figure}
\includegraphics[width=\columnwidth]{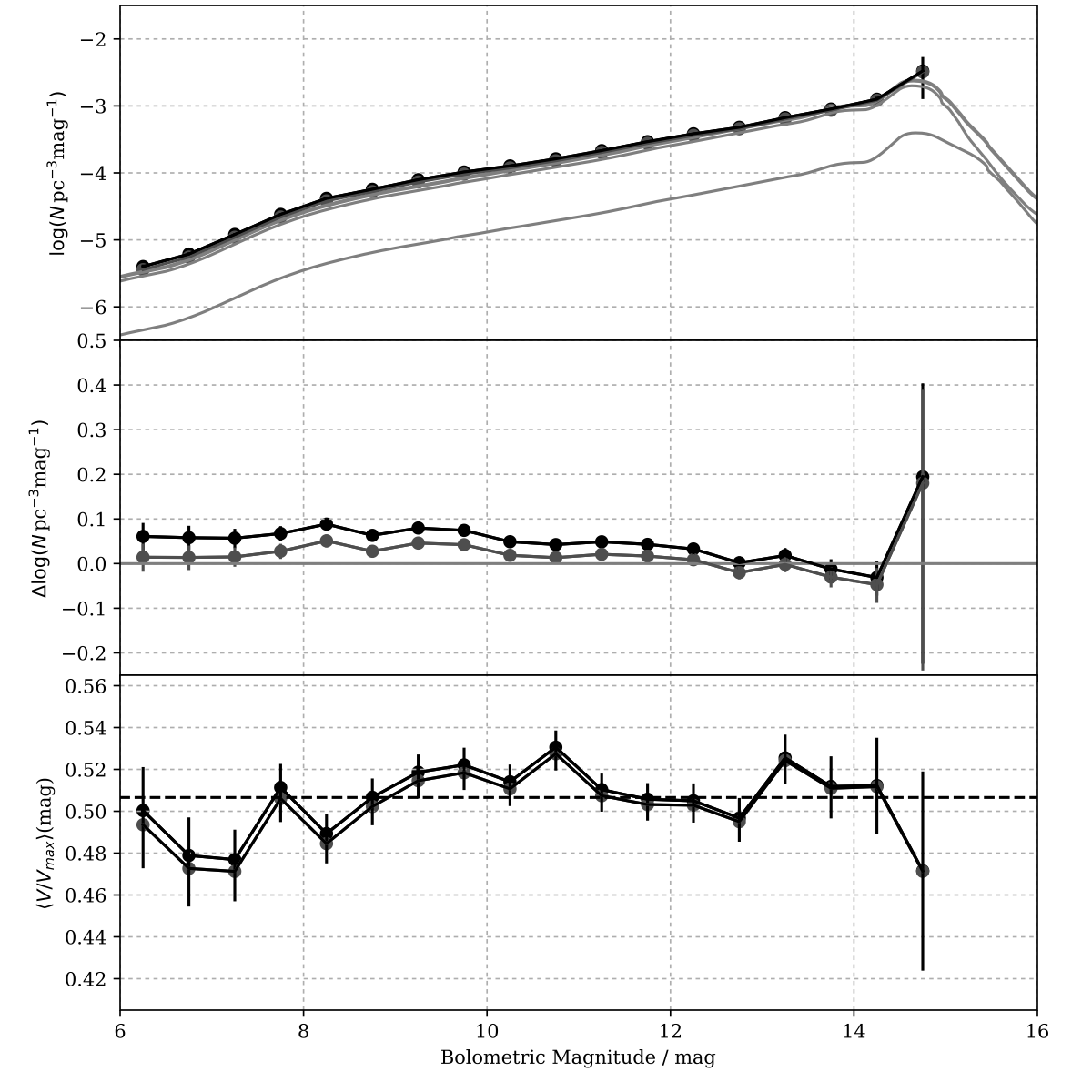}
\caption{Top: WLDFs for all sources with tangential velocity between $40$ and $120\kmps$ from a mixed-population catalogue using the new method assuming thin disc properties. Black uses a scale-height of $250$\,pc, grey uses $300$\,pc and the light grey line is the input function. Middle: the deviations of the WDLF as a function of magnitude. Bottom: the $\meanvvmax$ as a function of magnitude. The dashed line indicates the overall $\meanvvmax$.}
\label{fig:40_120_wdlf}
\end{figure}

In M17, $40$ and $120\kmps$ are used as the tangential velocity limits to study both discs together, with a scale-height of $300$\,pc instead of $250$\,pc. From H06, it is known that the effect of scale-height is larger at the bright end because sources can be seen at a larger distance hence the density-correction is larger in equation~(\ref{eq:vmax}). The effect on the total normalization is small because faint sources dominate after density and completeness corrections. However, studies in, for example, star formation history~\citep{2013MNRAS.434.1549R} or high energy exotic particles~\citep{2008ApJ...682L.109I} are sensitive to the whole range of magnitudes. This cannot simply be assumed to be negligible. Fig.~\ref{fig:40_120_wdlf} investigates this effect by comparing the cases of $250$ and $300$\,pc scaleheights. It shows that the choice of scale-height has almost no effect to the WDLFs except at the brightest magnitudes, and for the distribution of $\meanvvmax$, the differences are negligible. However, the absolute normalizations from using the two scale-heights are consistently overestimated by $\sim0.1$\,dex. In this simulation, $23.1\%$ of the sources in the range of $40-120\kmps$ are from the thick disc and the halo; in comparison, only $15.5\%$ of sources are not from the thin disc in the $40-60\kmps$ selection.

\subsection{Halo}

The study of the halo WDLF requires a selection of high velocity population in order to minimise the contaminations from the thick disc. In this section, we will show the observed WDLFs from a halo-only simulation and from a mixed thin disc, thick disc and halo simulation.

\subsubsection{Halo-only sample}

\begin{figure}
\includegraphics[width=\columnwidth]{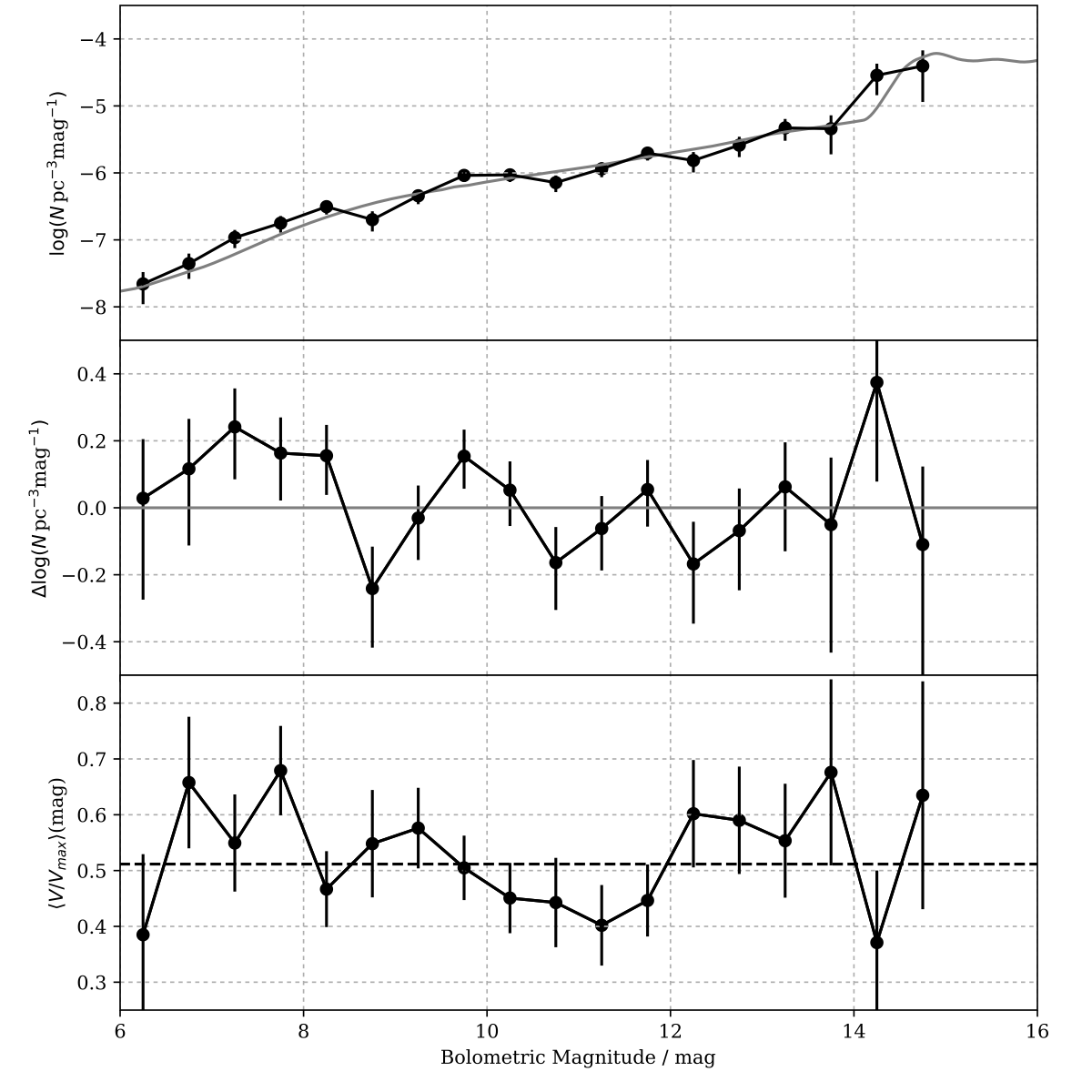}
\caption{Top: observed WLDFs~(black) for a halo-only realisation using the new method and the input function~(grey). Middle: the deviations of the WDLF as a function of absolute bolometric magnitude. Bottom: the $\meanvvmax$ as a function of absolute bolometric magnitude. The dashed line indicates the overall $\meanvvmax$.}
\label{fig:halo_wdlf}
\end{figure}

In the halo-only simulation, the observed WDLF agrees very well with the input function~(Fig.~\ref{fig:halo_wdlf}). From the $\meanvvmax(\mathrm{mag})$ distribution, the derived solution is stable throughout, except at the faintest bin where there are only two sources. The small oscillation about the line at $\meanvvmax=0.5$ is a good indication that the sample is unbiased over the entire range of magnitudes. The upper and lower tangential velocity limits are set at $200$ and $500\kmps$ which define the survey limits of the inner integral~(equations~\ref{eq:vmax} \& \ref{eq:inner_limit} and the equivalent set of the upper limit), the total integrated number density of the work is $1.77 \pm 0.10 \times 10^{-4}$ pc$^{-3}$, compared to the input $1.90 \times 10^{-4}$ pc$^{-3}$ (the integrated density up to $15\magnitude$ is $1.68 \times 10^{-4}$ pc$^{-3}$) and $\meanvvmax = 0.5116 \pm 0.0200$.

\subsubsection{Mixed Population~($200-500\kmps$)}

In 10 realizations of the halo-only simulation, under the selection of $200-500\kmps$, there is a mean contamination rate of $7.3\%$~(minimum at $3.6\%$ and maximum at $11.9\%$). In the thin disc analysis, where the contamination is over $15\%$, we do not observe any significant bias in the WDLF or $\meanvvmax$(mag) distribution. We believe these fractions of contaminations have little effect on the analysis so the samples should be representative of the halo, we therefore do not consider them further.

\subsection{Sensitivity to resolution}
\label{sec:resolution}
\begin{figure}
\includegraphics[width=\columnwidth]{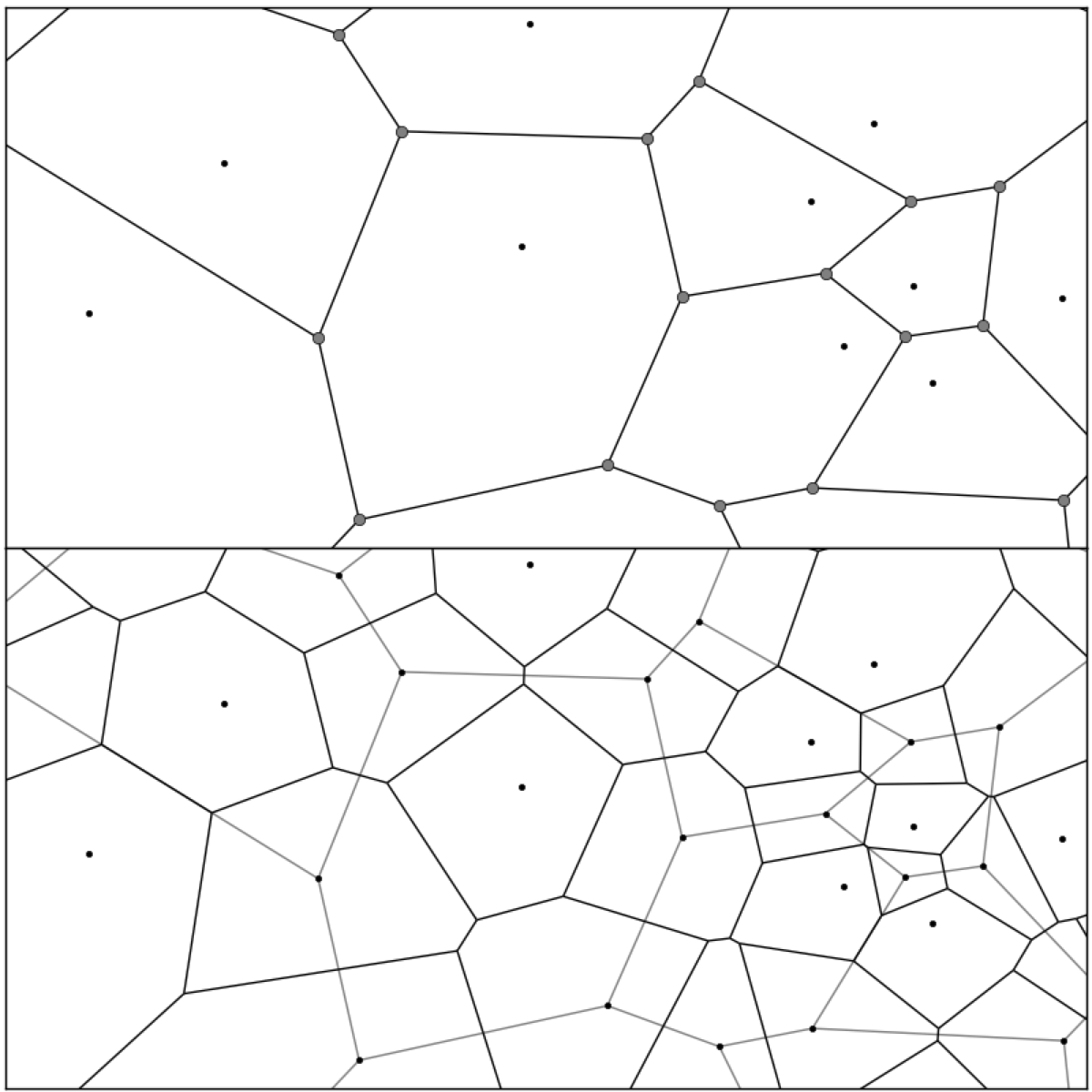}
\caption{Top: a Voronoi diagram with $15$ cells. The generating points~(only $10$ are shown) and the tessellation vertices, are displayed in black and grey respectively. Bottom: a Voronoi diagram with increased resolution generated with the points and vertices from the top panel, grey lines show the original cells.}
\label{fig:voronoi_resolution}
\end{figure}

\begin{figure}
\includegraphics[width=\columnwidth]{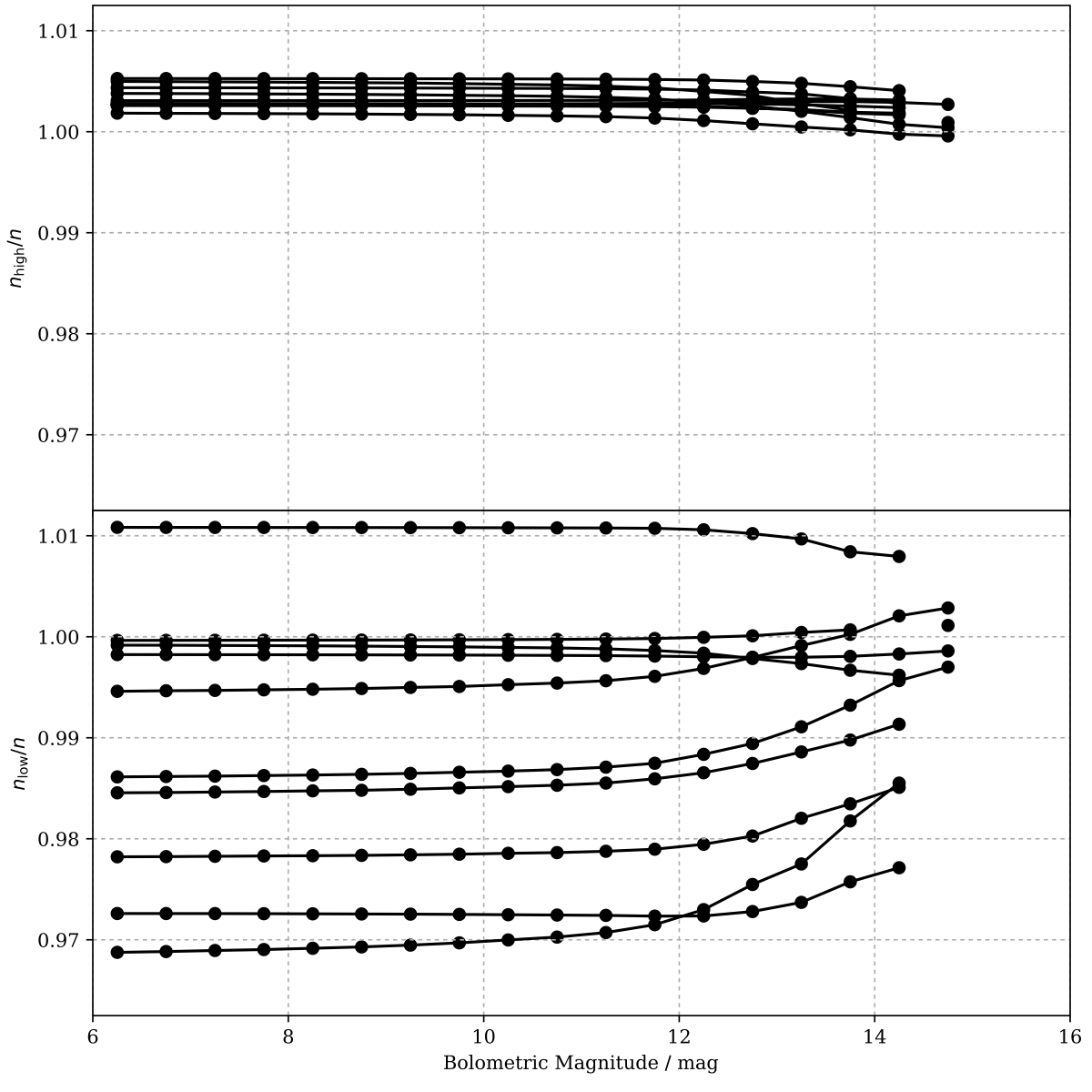}
\caption{Top: the ratio of number density between the high resolution~($\sim3\times$) analysis to the standard resolution as a function of absolute bolometric magnitude. Bottom: the ratio of number density between the low resolution~($\sim\frac{1}{3}\times$) analysis to the standard resolution as a function of absolute bolometric magnitude.}
\label{fig:resolution_compared}
\end{figure}

The purpose of this new method is to tackle a complex survey that has small-scale variations that reduce the maximum survey volume that is available to a study. The way this method divides the sky has avoided the detailed treatment of the survey and approximates it with the source properties that go into the analysis, which raises a concern whether the resolution for a small sample would be sufficient, and not causing significant systematic bias. We suggest a method to increase the resolution by a factor of $\sim3$, which can be repeated to further increase the resolution if needed, as it would be useful if there are only a few sources over the sky. A higher resolution can be achieved by using the vertices as new generating points~(Fig.~\ref{fig:voronoi_resolution}). However, an increased resolution requires much more computation time, because the time complexity of the Voronoi method is $\mathrm{O}\left( \mathcal{N}^2 \right)$. There is a trade-off between the accuracy and the computing time. The properties at the new cells can be approximated by those carried by the nearest sources~(or they can be extracted directly from the field from the raw data). Fig.~\ref{fig:resolution_compared} demonstrates with $10$ halo analyses that for a very well-behaved survey~(small differences in survey depths), a resolution in the order of $\sim30$\,deg$^{2}$ per cell compared to $\sim100$\,deg$^{2}$ per cell will only lead to an increase of $<1\%$ in number density~(top panel). The increase arises from the deeper parts of the survey that the standard resolution always underestimates~(as it is statistically less likely to land on the deeper parts). To understand the effect at lower resolutions, we simulate this by using one in three sources to generate the Voronoi tessellation. The change in number density is only in the range of a few percentage points~(bottom panel). Over a large number of simulations, the ratios should average to $1$. The asymmetric distribution comes from the inverse proportionality between the maximum volume and the number density.

\section{Conclusion}

In this work, we have demonstrated that the use of Voronoi tessellation can increase the survey volume to more optimally retrieve sources from a large sky area multi-epoch survey. The assumption behind this is not ideal, but it is not possible to take an average value of the number of epochs and their properties for each cell. Further subdivisions will take much more time to compute the volumes as this algorithm scales as $x \times \mathcal{N}^2$, where $x$ is the number of subdivision of each cell and $\mathcal{N}$ is the number of sources. Nevertheless, this method is one big step in approaching the optimal sampling of the survey footprint with limited computing power.

From a mixed population simulation, we find that under the framework of our galactic models, the new method recovers $\sim10-15\%$ more sources than the RH11 method under a typical lower proper motion selection. When considering a restricted tangential velocity selection~($40-60\kmps$), we do not observe any bias in the WDLFs brighter than $15\magnitude$. Similar result is observed from the $40-120\kmps$ sample. However, this conclusion is valid only up to $\mathrm{M}_{\mathrm{bol}}=15\magnitude$, the result should not be extrapolated to fainter magnitudes where thin disc contribution to the number density drops significantly compared to the thick disc and the halo. This work has deliberately removed sources closer than $100$\,pc to avoid bias caused by parallactic displacements, which as a consequence removed all the faint sources that can be seen only from a small distance. In the high velocity regime, at the given thick disc-to-halo density ratio, a $200-500\kmps$ selection will only contain a small fraction of contaminations so it is a good sample for studying the halo.

We have demonstrated one way to increase the resolution of the tessellation and it shows that for a well behaved survey, a low resolution only limits the volume by $\sim1\%$. An adaptive way that only subdivides cells larger than a certain solid angle can provide a grid of cells that have similar areas should it be more useful in some certain scenarios. When applying this method to real surveys, careful treatment at the boundaries is needed because the area of the survey is important. Leaving the boundaries untreated one will always end up with $4\pi$ steradian of sky area. In order to have a correct boundary that defines the survey, artificial points have to be added to create a layer of bounding cells surrounding the survey area such that the boundaries of the second last layer of cells overlap the survey footprint. One can identify the artificial points by using the survey boundary as a cell boundary, and then locate a generating point that can produce the correct boundary geometry. However, one should note that a typical survey boundary is given by a small circle on the celestial sphere, while the Voronoi tessellation constructs cell boundaries along the great circles. The total area described by the Voronoi cells is always going to be slightly different from the true survey area~(unless it is a full sky survey). However, the difference in area is very small, comparable to the unaccounted non-perfect fill-factor or dead pixels at the detector.

In future surveys, the complexity of the tiling pattern, scanning strategy as well as the detector arrays at the focal plane will only increase. There is an increasing need for a more optimal analytic tool to maximally use the available data. The Voronoi method can include the faintest objects that would otherwise be neglected because of unaccountable incompleteness.

\section*{Acknowledgments}
The Pan--STARRS 1 Surveys\,(PS1) have been made possible through contributions of the Institute for Astronomy, the University of Hawaii, the Pan--STARRS Project Office, the Max-Planck Society and its participating institutes, the Max Planck Institute for Astronomy, Heidelberg and the Max Planck Institute for Extraterrestrial Physics, Garching, The Johns Hopkins University, Durham University, the University of Edinburgh, Queen's University Belfast, the Harvard-Smithsonian Center for Astrophysics, the Las Cumbres Observatory Global Telescope Network Incorporated, the National Central University of Taiwan, the Space Telescope Science Institute, the National Aeronautics and Space Administration under Grant No. NNX08AR22G issued through the Planetary Science Division of the NASA Science Mission Directorate, the National Science Foundation under Grant No. AST-1238877, the University of Maryland, and Eotvos Lorand University\,(ELTE).

We thank the PS1 Builders and PS1 operations staff for construction and operation of the PS1 system and access to the data products provided. ML acknowledges financial support from the STFC Consolidated Grant of the Institute for Astronomy, University of Edinburgh. ML also wishes to thank Dr. Nigel Hambly and the referee Dr. Floor van Leeuwen for helpful comments and suggestions that have led to major improvements in the clarity and presentation of the article.

\bibliography{2017_Voronoi_Vmax}

\label{lastpage}

\end{document}